 \documentclass[final,3p,times,twocolumn]{elsarticle}

\usepackage{graphicx}

\usepackage{amssymb}

\usepackage{amsmath}

\journal{Astroparticle Physics}

\begin{document}

\begin{frontmatter}



\title{On the abundance of primordial bound states of superheavy magnetic monopoles}


\author{Dmitry\,R.~Gulevich}

\ead{D.R.Gulevich@lboro.ac.uk}
\address{Department of Physics, Loughborough University, Leicestershire, LE11 3TU, United Kingdom}


\begin{abstract}
It has been suggested that superheavy charged particles might have been born in primordial bound pairs at the end of cosmic inflation. Such pairs have been proposed as a source of ultra-high energy cosmic rays (UHECR).
We show that primordial bound pairs of magnetic monopoles larger than $10^{-9}$ cm 
quickly thermolise due to the interaction with primordial electron-positron plasma and any such initial primordial concentration is washed out. The final concentration will therefore be defined by their equilibrium abundance. 






\end{abstract}


\end{frontmatter}



\section{Introduction}

Superheavy magnetic monopoles arising as topologically stable solutions in Grand Unification Theories (GUT)~\cite{monopoles} might have been produced in the very early universe on the final stages of cosmic inflation. Monopoles of opposite charges would form bound states, monopoles-antimonopole pairs ($\rm{M\bar{M}}$ pairs) or "monopolonium"~\cite{Hill}. Contrary to its analogue positronium, monopolonium is formed by very heavy monopoles 
and behaves almost as a classical system of two particles orbiting their common centre of mass. Lifetime of a pair
can be estimated by the classical Larmor's formula for radiation of an accelerated charge. Assuming circular orbits one has in natural units~\cite{Hill, Dicus-Teplitz},
$\tau = M_m^{2} r^{3}/8 g^{4}$
where $r$ is separation of the particles, $M_m$ is mass of a monopole, and $g$ is magnetic charge.
For instance, a bound system of magnetic monopoles with $M_m\sim10^{16}$ GeV separated by $r=10^{-9}$ cm lives about $10^{18}$ s, of the order of the age of the Universe.
Due to the cubic dependence on $r$, lifetime of a pair with $r=10^{-10}$~cm is by 3 orders of magnitude smaller than in the former case.

Because of their long lifetime $\rm{M\bar{M}}$ bound states of monopoles formed in the early universe might have survived until present time and therefore are considered as a candidate for UHECR~\cite{Sigl,Huguet,pairs-2004,DVK-UHECR}.
Authors~\cite{Hill,Sigl} discussed $\rm{M\bar{M}}$ pairs formed in the equilibrium with free monopoles on the stage of nucleosynthesis.
Magnetic monopoles born in pairs at the end of the period of cosmic inflation have been discussed by~\cite{Turner, Trower} and recently by~\cite{pairs-2003, pairs-2004}.
The last authors~\cite{pairs-2003, pairs-2004} suggested that superheavy charged particles are born in non-equilibrium processes at the stage of preheating. 
Here we address the question of abundance of such primordial pairs made of magnetic monopoles.

The question of decay of magnetic monopole pairs has been discussed earlier in literature in the context of annihilation of massively overproduced free magnetic monopoles~\cite{Ya-B, Preskill,GKT, Dicus,Izawa,Preskill-1983} predicted by the GUT theories (the "monopole problem"). Once captured by the attractive forces, magnetic monopoles of opposite charges would form a bound pair which then quickly cascades down as described in~\cite{Preskill-1983}.
The arguments involve the drag force~\cite{GKT} acting on a monopole from the relativistic plasma which make a pair of monopoles to spiral down and finally annihilate~\cite{Blanco}. 
However, doubts  still remain in the scientific community~\cite{Dubrovich} if all of primordial magnetic monopole pairs 
were destroyed by this process.
%
Indeed, first, the use of classical relativistic mechanics for scattering of electrons on a magnetic monopole has to be justified, 
and second, existing calculations~\cite{Blanco} ignore the diffusive component due to the randomness of the momentum transfer to monopoles and are therefore deterministic.
Here we take into account the stochastic component in the evolution of $\rm{M\bar{M}}$ pairs and make use the available relativistic quantum cross sections.
We consider evolution of bound pairs of magnetic monopoles as a diffusion process in binding energy space of circular orbits.



\section{Stochastic evolution of monopole-antimonopole bound state}


Here we employ natural non-rationalized cgs-Gaussian system of electromagnetic units ($\hbar=c=k_B=1$). The fine structure constant is then $\alpha=e^2$ and Dirac charge quantization condition takes the form\footnote{In the Heaviside-Lorentz (rationalized) units the fine structure constant and the charged  quantization conditions are replaced by $\alpha=e^2/4\pi$ and $e\,g=2\pi n$.}
$e\,g=n/2$ with $n=\pm 1, \pm 2,...$.
Here we consider GUT monopoles of the minimal magnetic charge $g=1/2e\approx 5.85$ 
and mass $M_m \sim 10^{16}$ GeV.

Consider an $\rm{M\bar{M}}$ pair immersed into the sea of primordial relativistic electron-positron plasma at temperature $0.5\;\rm{MeV} \lesssim T \lesssim 100\;\rm{MeV}$. 
At $T\lesssim 100\;\rm{MeV}$ we may only consider contribution of electrons and positrons in the relativistic plasma.
After electron-positron annihilation at $T\lesssim m_e\approx 0.5\;\rm{MeV}$ the number of charged particles drops by many orders of magnitude and the dynamics of heavy monopoles is no longer influenced by the charged plasma. 





Due to high masses of magnetic monopoles a pair behaves as a purely classical system of bound particles orbiting its common centre of mass (at least, for the range of parameters of interest).
The two-body classical system of bound magnetic monopoles is reduced to one-body problem for a effective particle of reduced mass $\mu\equiv M_m/2$, position $\mathbf{r}=\mathbf{r}_1-\mathbf{r}_2$ and momentum $\mathbf{p}=(\mathbf{p}_1-\mathbf{p}_2)/2$. 
Assuming pairs on circular orbits (pairs on elliptical orbits have shorter lifetimes), the binding energy $\xi$ of $\rm{M\bar{M}}$ pair in terms of the magnitudes of $\mathbf{r}$ and $\mathbf{p}$ is
$$\frac{p^2}{2\mu}-\frac{g^{2}}{r}\equiv -\xi <0.$$
If $\mathbf{q}_1$, $\mathbf{q}_2$ are momenta transferred to each member of the pair in act of collision with a charged particle, the change in binding energy (more precisely, its absolute value) is
$$\Delta\xi = - \frac{\mathbf{p}\cdot\mathbf{q}}{\mu}-\frac{q^2}{2\mu}$$
where $\mathbf{q}=(\mathbf{q}_1-\mathbf{q}_2)/2$ and $q=|\mathbf{q}|$.
Evolution of a $\rm{M\bar{M}}$ pair immersed into the sea of electron-positron plasma can then be seen as a diffusion in the space of binding energy described by one-dimensional Focker-Planck equation~\cite{Risken} for distribution function $f(\xi)$,
\begin{equation}
\frac{\partial f}{\partial t}=\hat{L}_{FP} f
\label{FPeq}
\end{equation}
with the Focker-Planck operator
$$
\hat{L}_{FP}= -\frac{\partial}{\partial \xi}D^{(1)}(\xi) + \frac{\partial^2}{\partial \xi^2} D^{(2)}(\xi).
$$
where $D^{(n)}(\xi)$ are drift ($n=1$) and diffusion ($n=2$) coefficients~\cite{Risken}.
As the interaction time with a single particle is significantly smaller than the orbital period of the pair, the coefficients $D^{(n)}(\xi)$ can be evaluated by
\begin{multline}
D^{(n)}(\xi)=\frac{1}{n!}\lim_{\Delta t \to 0}\frac{\langle \left[\xi(t+\Delta t)-\xi(t)\right]^{n} \rangle}{\Delta t}\Big|_{\xi(t)=\xi}=\\
=\frac{1}{n!}\int \frac{d^{3}\mathbf{k}}{(2\pi)^{3}} \, f(\mathbf{k})v(k)\int d\sigma\,(\Delta\xi)^{n} 
\label{D1D2}
\end{multline}
where $\mathbf{k}$, $f(\mathbf{k})$, $v(k)$ and are momentum vector, distribution function and velocity of charged fermions (electrons and positrons), correspondingly. 









In the view of absence in the literature of 
relativistic quantum cross sections for scattering of electron on a bound magnetic monopole pair, we employ 
the impulse approximation~\cite{impulse-app} which allows us to use the available QED cross sections on a single monopole~\cite{Gamberg,cross-sections}.

The impulse approximation 
is applied when a single particle is incident upon a composite system consisting of two or more particles and can be summarized down to the three assumptions~\cite{Farina}: 
\par I. 
The range of interaction is small compared with the inter-particle distances. The incident particle interacts with only one particle of the target system during 
the collision. 
\par II. The target system can be regarded as transparent, so that the amplitude of the incident particle is not diminished in crossing the target system. 
\par III. The scattering occurs over such a short time that the effect of the binding 
forces during the collision may be neglected.

Assumption III is justified by the short collision times compared to the characteristic orbital time of a $\rm{M\bar{M}}$ pair. Assumptions I and II can be justified by considering a well separated pair.
Therefore, the range of momentum transfer due to scattering with plasma has to be limited from below by the inverse size of a $\rm{M\bar{M}}$ pair, $q \gg 1/r \equiv q_{min}$. On the other hand, the eikonal approximation 
valid in the low-momentum transfer regime, limits the transfer by the mass of the fermions
~\cite{Gamberg}, $q \ll m_e \equiv q_{max}$. From these arguments on can see that these analysis will work for sufficiently large pairs with $r\gg 1/m_e\approx 4\times 10^{-11}$ cm.







Due to the assumption I, 
the momentum transfer in a single collision with electron is therefore
$\mathbf{q}=\frac{1}{2}\mathbf{q}_1$, where $\mathbf{q}_1=\mathbf{k}-\mathbf{k}'$ is momentum transferred to one of the monopoles.
Introducing momentum transfer parallel $q_{1\parallel}$ and perpendicular $\mathbf{q}_{1\perp}$ to $\mathbf{k}$, one has in the low momentum transfer limit we have
$q_{1\parallel}={q_1^{2}}/{2k}$ and $q_{1\perp}\approx q_1$.
Therefore, we get for the change in the binding energy
\begin{equation}
\Delta\xi\approx - \frac{q^2}{2\mu}\left( 1 + 2\frac{p_\parallel}{k}\right) - \frac{p_\perp q \cos{\varphi}}{\mu}
\label{Dxi}
\end{equation}
where $\varphi$ is the angle between $\mathbf{p}_\perp$ and $\mathbf{q}_\perp$, $p_\parallel=p\cos{\theta_1}$, $p_\perp=p\sin{\theta_1}$ where $\theta_1$ is angle between momenta of a monopole and the scattered charged particle.

Relativistic differential cross section for scattering of electron on a single monopole calculated in the eikonal approximation~\cite{Gamberg} is (non-rationalized units)
${d\sigma}/{dq_1^{2}}=4\pi\,{(e g)^{2}}/{q_1^4}$
which also coincides with the classical calculation. 
With $q_1=2q$ and account of the second monopole as a factor of 2, the cross section for scattering on a bound state of monopoles is
$$\frac{d\sigma}{dq^{2}d\varphi}=\frac{(e g)^{2}}{q^4}$$

Distribution of charged particles of electron-positron plasma at temperature $T$ is described by the Fermi-Dirac distribution
which invloves chemical potential $\mu_e$.
At $T \gtrsim m_e$ number of electron-positron pairs is very large compared with the atomic electron density. 
Because the number electrons is almost equal to the number of positrons, with the sufficient accuracy their chemical potentials can be set to zero\footnote{see, e.g.~\cite{LL-V}, paragraph "Equlibrium with respect to pair production". More accurate calculations give $\mu_e/T\sim 10^{-9}$, see p.57 of Ref.~\cite{Lyth-Liddle}}.
As the monopoles are orbiting around their common centre of mass, there arises an anisotropic distribution of energies of scattered charged particles. In analogy with the dipole anisotropy of the Cosmic Microwave Background radiation~\cite{Weinberg}
we take this into account by introducing the effective anisotropic temperature $T(\theta_1)\approx T (1-v_1 \cos\theta_1)$ which depends on the orbital velocity $v_1\ll 1$ of a single monopole (see also Appendix A in~\cite{GKT}). 
We also disregard thermal motion of the c.m. of the pair through plasma and thus we take  isotropic  distribution of incoming particles in the rest frame of the $\rm{M\bar{M}}$ pair.
In the relativistic limit distribution of charged particles is then given by
$$
f(\mathbf{k})=\frac{g_{ch}}{e^{k/T(\theta_1)}+1}
$$
where $g_{ch}=4$ is the number of relativistic degrees of freedom of electrons and positrons. 
Evaluating the drift and diffusion coefficients~\eqref{D1D2}, and neglecting terms suppressed by mass of the monopole,
$$D^{(1)}(\xi)= -\alpha_0 + \alpha_1 \xi$$ 
$$D^{(2)}(\xi)= \beta \xi$$
with $\alpha_0, \alpha_1, \beta >0$ given by
$$\alpha_0=\frac{2\pi(eg)^{2}}{\mu}\, n_{ch}\ln \Lambda$$
$$\alpha_1=\frac{2\pi (eg)^{2} g_{ch}}{18\mu}\,T^2 \ln\Lambda$$
$$\beta=\frac{4\pi(eg)^{2}}{3\mu}\,n_{ch}\,\ln \Lambda$$
where $\ln\Lambda\equiv \ln(q_{max}/q_{min})$ and
$$
n_{ch}=\frac{3\zeta(3)\,g_{ch}}{4\pi^{2}}\,T^{3}
$$
is the number density of charged fermions.
Substitution of $f(t,\xi)=e^{-\lambda t}f(\xi)$ to~\eqref{FPeq} and neglecting the weak dependence of $\ln\Lambda$ on $\xi$, gives the equation on spectrum
$\hat{L}_{FP}f=-\lambda f$
which can then by change of variables $\xi=x\,\beta/\alpha_1$ be transformed to the associated Laguerre equation
$$x\, f''(x)+\left[1+\nu-x\right]f'(x)+\left(\frac{\lambda}{\alpha_1}-1\right)f=0$$
with $\nu=1+\alpha_0/\beta=5/2$, which possesses a discrete spectrum
$\lambda_n=\alpha_1 (n+1)$ for $n=0,1,2,...$
and non-singular solutions in the form of associated Laguerre polinomials $L_n^{(\nu)}(x)$~\cite{Suetin}.

Solution to the Focker-Planck equation is then represented as a series of associated Laguerre polinomials. If 
$$
f_0(\xi)=\sum_{n=0}^{\infty}c_n L_n^{(\nu)}(\alpha_1\xi/\beta)
$$
is the initial distribution\footnote{$f_0(\xi)$ can be an arbitrary function from $L_2[0,\infty;x^{\nu}e^{-x}]$ with $x=\xi\alpha_1/\beta$~\cite{Suetin}}
at certain time $t=0$, where coefficients $c_n$ are given by
$$
c_n=\frac{n!}{\Gamma(n+\nu+1)}\int_0^{\infty}x^{\nu}e^{-x} L_n^{(\nu)}(x) f_0(\beta x/\alpha_1) dx
$$
then at a later time
$$
f(\xi,t)=e^{-\alpha_1 t} \sum_{n=0}^{\infty}c_n e^{-n \alpha_1 t} L_n^{(\nu)}(\alpha_1\xi/\beta).
$$
As $L_0^{\nu}(\alpha_1\xi/\beta)=1$ and the higher order polynomials are suppressed by the exponential $\exp(-n\alpha_1 t)$, this results in flattening of the initial distribution with time, followed  by the exponential decay with characteristic relaxation time
$$\tau_{M}\equiv  \frac{1}{\alpha_1}=\frac{18}{4\pi (eg)^{2} g_{ch}\ln\Lambda}\,\frac{M_m}{T^2}$$
This time scale has to be compared with age of the Universe on the radiation dominated stage
$$
\tau_{RD}\equiv\frac{1}{2H}=\frac12 \left(g_{*}\frac{8\pi^{3}}{90}\right)^{-1/2}\frac{M_{pl}}{T^{2}}
$$
where $g_{*}$ is the number of relativistic degrees of freedom ($g_{*}=10.75$ for temperatures between 0.5 and 100 MeV).
Therefore,
$$
\frac{\tau_{M}}{\tau_{RD}}= \frac{1.3 \times 10^{-2}}{\ln\Lambda} \;\left(\frac{M_m}{10^{16}\;\rm{GeV}}\right)
$$ 
For $M_m=10^{16}$ GeV and $\ln\Lambda\gtrsim 3$ for pairs $r\gtrsim 10^{-9}$ cm, we have
$\tau_{M}/\tau_{RD} \lesssim 4\times 10^{-3}$ which is between one and two orders of magnitude larger than the  
estimate of the authors~\cite{Blanco} because of the mistake in their final formula\footnote{formula (14) in~\cite{Blanco} should have been 
$\tau_F/\tau_H\approx 4.5\; m_M/m_{pl}$ according to their calculation.}. Still, one gets a  discouraging upper limit on the abundance of primordial $M\bar{M}$ pairs
$$\frac{n^f_{M\bar{M}}}{n^i_{M\bar{M}}}\sim\exp(-\tau_{RD}/\tau_{M})\lesssim 10^{-100}$$ which implies that the abundance of magnetic monopole pairs after electron-positron annihilation $n^f_{M\bar{M}}$ is defined not by
the initial primordial concentration $n^i_{M\bar{M}}$ but by 
the thermal equilibrium with the background relativistic plasma and free monopoles.


\section{Conclusion}


Our analysis is applicable for the $\rm{M\bar{M}}$ pairs much larger than the electron Compton length $4\times 10^{-11}$ cm. Therefore, whereas we conclude that pairs larger than 
about $10^{-9}\;\rm cm$ 
are destroyed by the interaction with the primordial plasma before the electron-positron annihilation, we are not in a position to 
make a definite conclusion with respect to smaller pairs
for which our approximations 
are not reliable. 
Limited by the impulse approximation from below and by eikonal approximation from above, one should be discouraged pushing these two bounds too close to each other\footnote{Indeed, there are cases when impulse approximation fails to produce correct results as in the case of excitation of hydrogen by electron impact~\cite{Coleman-McDowell}.}. 
Coincidently, pairs $r\approx 10^{-9}\;\rm cm$ are those with lifetime of the order of age of our Universe and are on the edge of validity of our approximations. 
Therefore, we see two possible ways how this result can be improved.
In order to push this bound down to smaller pairs, either of the following cross sections must be calculated:
\par I. Cross section for scattering of relativistic electrons on a single magnetic monopole {\it beyond} the eikonal approximation, or,
\par II. Cross section for scattering of relativistic electrons on a {\it bound state} of magnetic monopoles.

If as a result, $\rm{M\bar{M}}$ pairs with lifespan of the age of our Universe turn out to be destroyed before the electron-positron annihilation stage, smaller pairs of magnetic monopoles, if survived, may still be 
of potential cosmological significance: as lifetime of $\rm{M\bar{M}}$ pairs with $r\approx 10^{-11}$ cm reaches the recombination epoch, they may potentially contribute
to the growth of primordial perturbations
or leave signature in the cosmic microwave background.


\section*{Acknowledgements}
The author would like to thank Victor Zalipaev and Alexei Vagov for support and valuable discussions and Victor Dubrovich for 
introducing to the problem.




\bibliographystyle{elsarticle-num}
\bibliography{<your-bib-database>}



\end{document}